\documentclass[12pt]{article}
\usepackage{bm}
\usepackage{amsmath}
\usepackage{amssymb}
\usepackage{float}
\usepackage{url}

\renewcommand{\thepage}{}
\makeatletter
\@addtoreset{equation}{section}
\renewcommand{\theequation}{\thesection.\@arabic\c@equation}
\makeatother
\renewcommand{\thefootnote}{\fnsymbol{footnote}}
\begin{document}
\begin{titlepage}
\title{
\vspace*{-4ex}
\hfill
\begin{minipage}{3.5cm}
\end{minipage}\\
\bf Gauge Invariant Overlaps for Identity-Based Marginal Solutions
\vspace{0.5em}
}

\author{Isao {\sc Kishimoto}$^{1}$
\ \ and\ \  
Tomohiko {\sc Takahashi}$^{2}$
\\
\vspace{0.5ex}\\
$^1${\it Faculty of Education, Niigata University,}\\
{\it Niigata 950-2181, Japan}\\
$^2${\it Department of Physics, Nara Women's University,}\\
{\it Nara 630-8506, Japan}}
\date{July, 2013}
\maketitle
%

\begin{abstract}
\normalsize
We investigate identity-based solutions associated with marginal
deformations in open string field theory. We find that the
identity-based marginal solutions can be represented as a difference
of wedge-based solutions plus an integration of a deformed BRST exact
state. Using this expression, the gauge invariant overlap can be
calculated analytically for the identity-based solutions.
Moreover, we show that, by gauge transformation, the overlap
is transformed into a disk correlation function with the integrations of
currents at the boundary.

\end{abstract}
\end{titlepage}

\renewcommand{\thepage}{\arabic{page}}
\renewcommand{\thefootnote}{\arabic{footnote}}
\setcounter{page}{1}
\setcounter{footnote}{0}
%
\section{Introduction}

Analytic classical solutions corresponding to the tachyon vacuum were
constructed in bosonic cubic open string field theory in a marginally
deformed background \cite{Inatomi:2012nv}.
The theory in the background is derived from
expanding a string field around the identity-based marginal
solution \cite{Takahashi:2001pp,Takahashi:2002ez,Kishimoto:2005bs}, 
which is an analytic solution in the original theory characterized by
the BRST operator $Q_B$.
A remarkable feature of the tachyon vacuum solution in the background is
that the vacuum energy of the solution can be calculated exactly,
although that of the
identity-based marginal solution is given as an indefinite quantity.
The resulting vacuum energy of the tachyon vacuum solution
allows us to expect that the vacuum energy
of the identity-based solution vanishes.

The gauge invariant
overlap \cite{Zwiebach:1992bw,Hashimoto:2001sm,Gaiotto:2001ji} can also
be evaluated analytically for the tachyon vacuum solution in the
marginally deformed background \cite{Inatomi:2012nv}. It is defined as,
for a closed string vertex operator $V(i)$ on the midpoint, 
\begin{align}
 O_V(\Psi)&=\left<I|V(i)|\Psi\right>,
\label{overlap}
\end{align}
where $\Psi$ is an open string field and $I$ is the identity string
field. The resulting overlap for the tachyon vacuum solutions reproduces
correctly the 
effect of marginal deformations for open-closed string couplings.
However, as in the case of the vacuum energy, the overlap for the
identity-based marginal solution is apparently indefinite.

These observables for the identity-based solution become
indeterminate
because a string field is a state in a Hilbert space with an indefinite
metric. For instance, if expanding the solution in a Fock space, the
observables are indeterminate forms of type $\infty-\infty$.
In general, the indefiniteness occurs widely in string field theory
and several possible ways are suggested to avoid
divergence~\cite{Yoneya:1987gc,Kawano:1992dp,Kugo:1992md,Bonora:2011ri,
Bonora:2011ru,Bonora:2011ns,Bonora:2010hi,Bonora:2013cya,
Erler:2011tc,Erler:2012dz,Murata:2011ep,Hata:2011ke}. 
To obtain a definite value, it is necessary that some sort of
regularization technique makes the observables finite; this is probably
controlled by symmetry.

The purpose of this paper is to calculate
the gauge invariant overlaps analytically for the identity-based
marginal solution. By the definition (\ref{overlap}), the overlap is
linear with respect to a string field, in contrast to the vacuum
energy. In addition, it is invariant under BRST transformation in any
background. In fact, we find that 
\begin{align}
 O_V(Q_\phi \Lambda)&=0,
\label{OVQ}
\\
 Q_\phi \Lambda&= Q_B\Lambda+\phi*\Lambda-(-1)^{|\Lambda|}\Lambda*\phi,
\label{deformedBRS}
\end{align}
where $Q_\phi$ is a deformed BRST operator in the background generated
by a classical solution $\phi$ in the original theory.
These properties have an essential role in the analytic calculation of the
overlap for the identity-based solution.
We will find that, owing to the linearity and the BRST invariance,
a finite value of the overlap can be derived by a norm-cancellation
mechanism.

The level truncation analysis works well in
the theory expanded around identity-based
solutions~\cite{Takahashi:2003ppa,Kishimoto:2009nd,Kishimoto:2011zza}
and permits us to check exact results indirectly.
Actually, numerical solutions have been constructed in
a marginally deformed background~\cite{Kishimoto:2013dna}.
We note that numerical
overlaps for the 
solutions consistently support our analytic results given for the
identity-based marginal solutions.

This paper is organized as follows. In section \ref{sec:analytic}, 
we briefly review the
identity-based marginal solution in the original theory and the tachyon
vacuum solution in the deformed background.
Next, we provide a new expression for the identity-based marginal
solution in terms of the tachyon vacuum solutions. Using this
expression, we calculate the overlap for the identity-based solution.
Then, we discuss the idea that the resulting overlap includes an
expression that has been
given for wedge-based marginal solutions.
In section \ref{sec:RM}, we give some concluding remarks.
In Appendix A, we give an expression of the overlap for the tachyon
vacuum solutions.

\section{Analytic evaluation of the gauge invariant overlap for
  identity-based marginal solutions
\label{sec:analytic}
}

\subsection{Identity-based marginal solutions and tachyon vacuum
  solutions} 

We briefly summarize identity-based marginal solutions in the
original theory characterized by the BRST charge $Q_B$ and a tachyon
vacuum solution in the theory expanded around the marginal solutions.

The action in bosonic open cubic string field theory is given by
\begin{align}
 S[\Psi;\,Q_B]&=\int \left(\frac{1}{2}\Psi*Q_B\Psi
+\frac{1}{3}\Psi*\Psi*\Psi\right),
\end{align}
where $Q_B$ is the original BRST charge.  The equation of motion is
$Q_B\Psi+\Psi*\Psi=0$. Using holomorphic current operators $j^a(z)$
associated with a Lie algebra, and the 
identity string field $I$, we have an identity-based marginal
solution~\cite{Kishimoto:2005bs}:
\begin{align}
 \Psi_0&=-V_L^a(F_a)I-\frac{1}{4}g^{ab}C_L(F_aF_b)I.
\label{msol}
\end{align}
The half-string operators are defined by
\begin{align}
&V_L^a(f)=\int_{C_{\rm left}}\frac{dz}{2\pi
 i}\frac{1}{\sqrt{2}}f(z)\,c\,j^a(z),
~~~~~~
C_L(f)=\int_{C_{\rm left}}\frac{dz}{2\pi i}f(z)\,c(z),
\end{align}
where $c(z)$ is the ghost operator and $f(z)$ is a function on the unit
circle $|z|=1$. 
$C_{\rm left}$ is the path along a unit half circle: ${\rm
Re}\,z\ge 0$.
The function $F_a(z)$ in (\ref{msol}) satisfies
$F_a(-1/z)=z^2F_a(z)$. The parameters describing marginal deformations
are given by
\begin{align}
 f_a&=\int_{C_{\rm left}}\frac{dz}{2\pi i}F_a(z).
\label{mpara}
\end{align} 

Expanding the string field around the solution (\ref{msol}) as
$\Psi=\Psi_0+\Phi$, we obtain the action in a marginally 
deformed background: $S[\Phi;\,Q_{\Psi_0}]$.
Here, the deformed BRST operator $Q_{\Psi_0}$ is given by the definition
(\ref{deformedBRS}). 
As a feature of the identity-based solution,
we have a simpler expression for the deformed BRST operator:
\begin{align}
 &Q_{\Psi_0}= Q_B-V^a(F_a)-\frac{1}{4}g^{ab}C(F_aF_b),
\label{deformedBRSTm}
\\
 &V^a(f)=\oint \frac{dz}{2\pi i}\frac{1}{\sqrt{2}}f(z)\,cj^a(z),
~~~~~~~ 
C(f)=\oint \frac{dz}{2\pi i}f(z)\,c(z),
\end{align}
where the integration path is given by a whole unit circle.

The equation of motion in the deformed background is given by
$Q_{\Psi_0}\Phi+\Phi*\Phi=0$. Then, we can construct a tachyon vacuum
solution in the marginally deformed background \cite{Inatomi:2012nv}:
\begin{align}
 \Phi_T&=\frac{1}{\sqrt{1+K'}}(c+cK'Bc)\frac{1}{\sqrt{1+K'}}.
\label{tsol}
\end{align}
Here, $K'$ is defined by $K'=Q_{\Psi_0}B$ and therefore it is given as
a state deformed from $K\,(=Q_BB)$ by the currents and a numerical
term:
\begin{align}
 K'&=K+J,\\
 J&= 
-\frac{\pi}{2}\int_{C_{\rm left}}\frac{dz}{2\pi i}
\frac{1}{\sqrt{2}}(1+z^2)F_a(z)j^a(z)I
-\frac{\pi}{8}\int_{C_{\rm left}}\frac{dz}{2\pi i}(1+z^2)g^{ab}
F_a(z)F_b(z)I.
\end{align}
$K'$, $B$, $c$ and $Q_{\Psi_0}$ have the same algebraic
structure as that of the $KBc$ algebra with $Q_B$ and, in analogy with
the Erler-Schnabl solution \cite{Erler:2009uj},
we can easily find that $\Phi_T$ is a classical solution in the
marginally deformed background.

Further expanding the string field around the solution (\ref{tsol}), we
find the kinetic operator $Q'_{\Phi_T}$ at the tachyon vacuum.
For an arbitrary string field $\Xi$, $Q'_{\Phi_T}$ is defined as
\begin{align}
 Q'_{\Phi_T}\Xi&=Q_{\Psi_0}\Xi+\Phi_T*\Xi-(-1)^{|\Xi|}\Xi*\Phi_T.
\end{align}
Thanks to the $K'Bc$ algebra, we can construct a homotopy operator
$\hat{A}$ such that $\{Q'_{\Phi_T},\hat{A}\}=1$ and $\hat{A}^2=0$:
\begin{align}
&\hat{A}\,\Xi=\frac{1}{2}(A*\Xi+(-1)^{|\Xi|}\Xi*A),
&A=\frac{1}{\sqrt{1+K'}}B\frac{1}{\sqrt{1+K'}}.
\label{homotopy}
\end{align}
Therefore, we can formally prove that $Q'_{\Phi_T}$ has vanishing
cohomology.

\subsection{The gauge invariant overlaps for identity-based marginal
  solutions
\label{sec:2.2}
}

Let us introduce a parameter $t$ into the weighting functions
of the solution (\ref{msol}) as $F_a^t(z)=t F_a(z)$. We denote the
solution given by $F_a^t(z)$ as $\Psi_0^t$:
\begin{align}
 \Psi_0^t&=-t\,V_L^a(F_a)I-\frac{t^2}{4}g^{ab}C_L(F_aF_b)I.
\end{align}
$\Psi_0^t$ provides a one-parameter family of identity-based solutions
connecting a trivial configuration and $\Psi_0$.

Correspondingly, we can consider the tachyon vacuum solution
$\Phi_T^t$, 
which is constructed by using $K'_t=Q_{\Psi_0^t}B$ instead of $K'$ in
$\Phi_T$ (\ref{tsol}). 
$\Phi_T^t$ becomes $\Phi_T$ for $t=1$ and the Erler-Schnabl
solution $\Phi_T^{\rm ES}$
for $t=0$.\footnote{The Erler-Schnabl solution is given by
$\Phi_T^{\rm ES}=\frac{1}{\sqrt{1+K}}(c+cKBc)\frac{1}{\sqrt{1+K}}$.} 
Similarly, we can find that 
a homotopy operator exists for $Q'_{\Phi_T^t}$,
which is given by replacing $K'$ in (\ref{homotopy}) with
$K'_t$.

Now, let us consider the sum of the identity-based marginal solution
$\Psi_0^t$ and the tachyon vacuum solution $\Phi_T^t$ in the deformed
background:
$\Psi_T^t=\Psi_0^t+\Phi_T^t$.
It connects $\Psi_T^{t=0}=\Phi_T^{\rm ES}$ and $\Psi_T^{t=1}=\Psi_0+\Phi_T$.
We find that $\Psi_T^t$ is a classical solution in
the original background. In fact, by adding the two equations,
$Q_B\Psi_0^t+\Psi_0^t*\Psi_0^t=0$ and
$Q_{\Psi_0^t}\Phi_T^t+\Phi_T^t*\Phi_T^t=0$, we 
have the equation,
\begin{align}
 &Q_B\Psi_T^t+\Psi_T^t*\Psi_T^t=0.
\label{eom}
\end{align}

Expanding the string field around $\Psi_T^t$ in the original theory, 
we have the deformed BRST operator $Q_{\Psi_T^t}$ by following the
definition (\ref{deformedBRS}). 
Substituting $\Psi_0^t+\Phi_T^t$ for $\Psi_T^t$ in
$Q_{\Psi_T^t}$, we
can easily find that $Q_{\Psi_T^t}$ is the kinetic operator at the
tachyon vacuum in the deformed background; i.e.,
$Q_{\Psi_T^t}=Q'_{\Phi_T^t}$.
We remember that $Q'_{\Phi_T^t}$ has no cohomology
because of the existence of a homotopy operator as mentioned
above.
Consequently, we conclude that $Q_{\Psi_T^t}$-closed states are
$Q_{\Psi_T^t}$-exact states.

Differentiating (\ref{eom}) with respect to $t$, we have
\begin{align}
 Q_{\Psi_T^t}\frac{d}{dt}\Psi_T^t&=0.
\end{align}
Then, we find that, for some state $\Lambda_t$,\footnote{
We can take $\Lambda_t=\hat{A}^t\frac{d}{dt}\Psi_T^t$
using the homotopy operator $\hat{A}^t$.}
\begin{align}
\frac{d}{dt}\Psi_T^t = Q_{\Psi_T^t}
\Lambda_t.
\label{dtPsit}
\end{align}
Integrating (\ref{dtPsit}) from $0$ to $1$, we get
\begin{align}
 \Psi_0&= \Phi_T^{\rm ES}-\Phi_T +\int_0^1 Q_{\Psi_T^t}
\Lambda_t\,dt.
\label{idm}
\end{align}
Thus, the identity-based marginal solution $\Psi_0$ is expressed by
the difference of the wedge-based solutions plus the integration of a
deformed BRST exact state.

Equation (\ref{idm}) is a useful expression
for evaluating the gauge invariant overlaps
for the identity-based marginal solution. 
Substituting (\ref{idm}) into (\ref{overlap}), 
it can be written as
\begin{align}
 O_V(\Psi_0)&=O_V(\Phi_T^{\rm ES})-O_V(\Phi_T),
\label{OVPsi}
\end{align}
where the contribution of the last term in (\ref{idm}) vanishes
due to the BRST invariance (\ref{OVQ}). 
Although the left hand side of (\ref{OVPsi}) seems to be
indefinite due to singular property of the identity string field,
the right-hand side can be calculated analytically.
In fact, $O_V(\Phi_T^{\rm ES})$ and $O_V(\Phi_T)$ have been already
computed in Refs.~\cite{Erler:2009uj} and \cite{Inatomi:2012nv},
respectively.
Thus, the overlap for the identity-based marginal solution is reduced to
the difference of the overlaps for wedge-based solutions.

As a first concrete example, let us consider the overlap with the
graviton vertex operator, 
$V\sim c\bar{c}\partial X^0 \bar{\partial} X^0$ and the identity-based solution
generated by spatial currents. Since $V$ does not correlate with spatial
currents, we find the relation $O_{V}(\Phi_T)=O_{V}(\Phi_T^{\rm ES})$ as
discussed in Ref.~\cite{Inatomi:2012nv}. From the relation (\ref{OVPsi}), it
follows that the overlap for the identity-based marginal solution is
zero: $O_V(\Psi_0)=0$.

Secondly, let us consider the identity-based marginal
solution generated by the current
\begin{align}
 j(z)&=\frac{i}{\sqrt{2\alpha'}}\partial X^{25}(z).
\label{j25}
\end{align}
Then, we consider the case that the 25th direction is compactified on a
circle of radius $R$ and the vertex operator $V$ is given as
\begin{align}
 V&=\tilde{V} e^{ik_L X^{25}(i)+ik_R X^{25}(-i)},
\end{align}
where $\tilde{V}$ is an operator containing no $X^{25}$.
In Ref.~\cite{Inatomi:2012nv}, it is shown that a phase shift occurs in the
overlap when the background is deformed by the current (\ref{j25}):
\begin{align}
 O_V(\Phi_T)&=\exp\left(i\frac{\pi
wR}{\sqrt{\alpha'}}f\right)O_V(\Phi_T^{\rm ES}),
\end{align}
where $w$ is the winding number in the $25$th direction\footnote{
$w$ is related to $k_L$ and $k_R$ as $w=(k_L-k_R)\alpha'/R$.}
and $f$ is a marginal parameter given by
Eq.~(\ref{mpara}) \footnote{Since we consider an Abelian case,
$f$ is given by using a function $F(z)$: $f=\int_{C_{\rm
left}}\frac{dz}{2\pi i}F(z)$.} for the current (\ref{j25}).

Consequently, from (\ref{OVPsi}),
we obtain
\begin{align}
 O_V(\Psi_0)&=
\left\{1-\exp\left(i\frac{\pi wR}{\sqrt{\alpha'}}f\right)\right\}
 O_V(\Phi_T^{\rm ES}).
\label{OVPsiex}
\end{align}
It should be noted that this is a non-zero
analytic result for physical observables of the identity-based solutions.

\subsection{The relation to one-point functions}

The gauge invariant overlap for $\Phi_T$ (\ref{tsol})
 is represented by using a correlation function on a cylinder
of circumference $\pi$ (or on the sliver frame obtained by 
the conformal mapping $\arctan z$ from the canonical upper half plane)
 \cite{Inatomi:2012nv, Inatomi:2012nd}:
\begin{align}
 O_V(\Phi_T)=\frac{e^{-\pi{\cal C}}}{\pi}
\left\langle V(i\infty)c\left(\frac{\pi}{2}\right)
\exp\left(\int_{-\frac{\pi}{2}}^{\frac{\pi}{2}}
 dx{\cal J}(x)\right)\right\rangle_{C_{\pi}}
\label{correlfunc}
\end{align}
where ${\cal J}$ is given in terms of the current $j^a$ 
in the marginal solution (\ref{msol}), 
and ${\cal C}$ comes from their operator product expansion (OPE):
\begin{align}
{\cal J}(x) &= \int_{-\infty}^{\infty}dyf_a(y)j^a(x+iy),
\label{calJt}
\\
{\cal C}&=\int_{-\infty}^{\infty}dy(-\pi) g^{ab}f_a(y)f_b(y).
\label{calCdef}
\end{align}
In the above integrations, the weighting function $f_a(y)$ is
given by
\begin{align}
 f_a(y)&= \frac{F_a(\tan(iy+\frac{\pi}{4}))}{2\pi\sqrt{2}
\cos^2(iy+\frac{\pi}{4})}.
\label{f_ay}
\end{align}
We give a derivation of (\ref{correlfunc}) in Appendix \ref{sec:App}.
In the correlation function (\ref{correlfunc}), the current operator is
integrated inside the cylinder, which is parametrized by $x+iy$.
In contrast, the gauge invariant overlap 
for the wedge-based solutions \cite{Schnabl:2007az, Kiermaier:2007ba,
Fuchs:2007yy, Kiermaier:2007vu}
is expressed by a correlation function with the $x$-integrations of
currents inserted at the open string boundary, namely $y=0$
\cite{Ellwood:2008jh, Kishimoto:2008zj}. 
The boundary insertion is a natural consequence from the viewpoint of
boundary marginal
deformations in conformal field theory. 
However, this is not a necessity in string field theory.

Here, we will show that the expression (\ref{correlfunc}) can be
related to a one-point function with the boundary insertion. 
First, let us take $F_a(z)$ in $\Psi_0$ as
\begin{align}
F_a(z;\,s) &=
 2\lambda_a\,\frac{s(1-s^2)}{\arctan\left(\frac{2s}{1-s^2}\right)}
\frac{\left(z+\frac{1}{z}\right)\frac{1}{z}}{
1-s^2\left(z^2+\frac{1}{z^2}\right)+s^4},
\label{F_a_def}
\end{align}
where $s$ is a parameter from $0$ to $1$ and $\lambda_a$ corresponds to
a marginal parameter by following (\ref{mpara}):
\begin{align}
 \int_{C_{\rm left}}\frac{dz}{2\pi i}F_a(z;\,s)&=\frac{2\lambda_a}{\pi}.
\end{align}
This indicates that the marginal parameters are 
independent of the parameter $s$ in (\ref{F_a_def}).
Since the form of $F_a$ except the half integration mode (\ref{mpara})
can be changed by gauge transformations of $\Psi_0$
\cite{Kishimoto:2005bs}, the parameter $s$ is redundant in the
gauge invariant overlap (\ref{OVPsi}) and therefore in (\ref{correlfunc}).

For the limit $s\rightarrow 0$, the function becomes
\begin{align}
 F_a(z;\,s)&\rightarrow \lambda_a\left(z+\frac{1}{z}\right)\frac{1}{z}.
\end{align}
This limiting form is a function discussed as a simple
example in Ref.~\cite{Inatomi:2012nv}.  

Taking the limit $s\rightarrow 1$, the function approaches the
delta function. In fact, (\ref{F_a_def}) can be expanded into a Laurent
series:
\begin{align}
 F_a(z;\,s)&=2\lambda_a \frac{1}{\arctan\left(\frac{2s}{1-s^2}\right)}
\sum_{n=0}^\infty
 s^{2n+1}\left(z^{2n+1}+\frac{1}{z^{2n+1}}\right)\frac{1}{z}.
\end{align}
Then, putting $z=e^{i\theta}$, we find that, for $s\rightarrow 1$, 
\begin{align}
F_a(z;\,s)&\rightarrow
4\lambda_a \frac{1}{\pi}\sum_{n=0}^\infty
 \left(z^{2n+1}+\frac{1}{z^{2n+1}}\right)
\frac{1}{z}
=
4\lambda_a \left\{\delta(\theta)+\delta(\pi-\theta)\right\}.
\end{align}
On the unit disk $|z|=1$, $\theta=0$ and $\theta=\pi$ correspond to open
string boundaries. As a result,
the support of the function $F_a(z;\,s)$ localizes
at the open string boundaries for the limit.

In the sliver frame, the line $u=\pi/4+i y$ corresponds to
the left half of an open string, $|\theta|\leq \pi/2$. Following the
relation $e^{i\theta}=\arctan(\pi/4+iy)$, we find that
$\theta$ is given by a function of $y$:
\begin{align}
 \theta(y)&=\arctan\left(\sinh 2y\right).
\end{align}
Then, for the limit $s\to 1$, 
the function $f_a(y)$ (\ref{f_ay}) approaches
$\frac{\sqrt{2}}{\pi}\lambda_a\delta(y)$ and
then the integration ${\cal J}(x)$
(\ref{calJt}) becomes a local operator:
\begin{align}
 {\cal J}(x)&\rightarrow \frac{\sqrt{2}}{\pi}\lambda_a\,j^a(x).
\end{align}
Consequently, in the limit $s\to 1$,
(\ref{correlfunc}) can be expressed as
the correlation function deformed by the boundary operator:
\begin{align}
\frac{e^{-\pi{\cal C}}}{\pi}
\left\langle\!V(i\infty)c(\frac{\pi}{2})
\exp\!\left(\int_{-\frac{\pi}{2}}^{\frac{\pi}{2}}
 \!dx {\cal J}(x)\right)\!\right\rangle_{C_{\pi}}
 &\rightarrow 
\frac{e^{-\pi{\cal C}}}{\pi}
\left\langle\! V(i\infty)c(\frac{\pi}{2})
\exp\!\left(\frac{\sqrt{2}}{\pi}\lambda_a\!\int_{-\frac{\pi}{2}}^{\frac{\pi}{2}}
 \!dx j^a(x)\right)\!\right\rangle_{C_{\pi}}.
\label{correlfunc2}
\end{align}
In general, the boundary integration in (\ref{correlfunc2})
diverges whenever the OPE among the currents
contains poles. However, the correlation function
(\ref{correlfunc}) does not suffer from divergence \cite{Inatomi:2012nv}
and then, as far as $s\neq 1$, (\ref{correlfunc2}) can be calculated
finitely. 
Accordingly, it turns out that the divergence in the limit 
should be canceled by the factor depending on ${\cal C}$
(\ref{calCdef}),
which diverges in the limit $s\to 1$.
In other words, the function $F_a(z; s)$ can be applied to regularize
the boundary integration of the currents and then ${\cal C}$ takes a
role as a subtraction term to cancel the divergence.\footnote{
It should be noted that this is
essentially the same result as that found by T.~Erler and C.~Maccaferri,
which was presented at the SFT2012 conference in Jerusalem.
According to their discussion, the gauge invariant overlap can be
computed by using the ``phantom term'' \cite{Erler:2012qr,Erler:2012qn},
which emerges when connecting the two solutions, 
the solution $\Psi_0$ and 
the tachyon vacuum solution $c(1-K)$. Taking the limit in
the phantom term, the bulk
integration of the current in the overlap is localized at the
open string boundary. Like our results, ${\cal C}$ serves as a
subtraction term to cancel the divergence.
Consequently, the correlation function can be regarded as the
disk amplitude with the integration of the currents at the
boundary.}

Finally, from the above result of $O_V(\Phi_T)$ and the relation
(\ref{OVPsi}),
we conclude that the overlap of the identity-based marginal solution is
represented by the difference between the two disk amplitudes:
\begin{align}
 O_V(\Psi_0)&=
 \frac{1}{\pi}\left<V(i\infty)c(\frac{\pi}{2})
\left\{1-
e^{-\pi{\cal C}}
\exp\left({\frac{\sqrt{2}}{\pi}\lambda_a\,\oint dx\,j^a(x)}
\right\}\right)
\right>_{C_{\pi}}.
\end{align}
This expression just corresponds to the result
in Ref.~\cite{Ellwood:2008jh} for wedge-based marginal
solutions.

\section{Concluding remarks
\label{sec:RM}
}

We have shown that the gauge invariant overlaps for identity-based
marginal solutions can be calculated exactly
in terms of the overlaps for tachyon vacuum solutions.
We emphasize that the relation (\ref{idm}) enables us to derive 
non-zero values of the overlaps for identity-based solutions.
Then, we have found that, by gauge transformations, the resulting
overlap is transformed to the disk amplitude with the boundary
deformation.
The result is identical to the gauge invariant overlap for
wedge-based marginal solutions.

The relation (\ref{idm}) is useful for evaluating the gauge invariant
overlaps for the identity-based solutions. In addition,
the linearity and the BRST invariance of the overlaps are important in
the calculation.
However, it remains difficult to calculate the vacuum energy of the
identity-based solutions directly even if (\ref{idm}) is used, since the
vacuum energy is non-linear with respect to the string field.

In \S \ref{sec:2.2}, by using (\ref{idm}), the gauge invariant
overlap with the
graviton vertex $V\sim c\bar{c}\partial X^0\bar{\partial} X^0$
is found to be zero for the identity-based marginal solutions.
Therefore, we can expect that the vacuum energy vanishes 
if we use the proportionality between the energy and the overlap
with the vertex \cite{Baba:2012cs}.
This is consistent with the result that was previously
evaluated by the differentiation of the vacuum energy with respect to
the marginal parameters
\cite{Takahashi:2001pp,Takahashi:2002ez,Kishimoto:2005bs}.

Our non-trivial analytic results are seen as a significant step in
understanding identity-based solutions. The results seem to be easily 
extended to the case of identity-based marginal solutions in superstring
field theories \cite{Kishimoto:2005bs,Kishimoto:2005wv}.
One of the most important applications is the
evaluation of the overlap and the vacuum energy for the identity-based
tachyon vacuum solution
\cite{Takahashi:2002ez,Kishimoto:2009nd,Kishimoto:2002xi}.
Although it
is not straightforward to apply our 
results to the tachyon vacuum case, we expect that they offer a new
perspective on identity-based solutions.

\section*{Acknowledgements}
We are grateful to Ted Erler and Carlo Maccaferri for helpful
discussions.
We would like to thank Hiroyuki Hata for his encouragement.
The work of I. K. and T. T. is supported by a JSPS Grant-in-Aid for
Scientific Research (B) (\#24340051). 
The work of I. K. is supported in part by a
JSPS Grant-in-Aid for Young Scientists (B) (\#25800134).

\appendix

\section{A derivation of (\ref{correlfunc})
\label{sec:App}}

In the marginally deformed background \cite{Inatomi:2012nv}, the
overlaps for the tachyon vacuum solution are given by
\begin{align}
 O_V(\Phi_T)&={\rm Tr}\left(V c \frac{1}{1+K'}\right)
 =\int_0^\infty dt\,e^{-t}\,
\langle
 I|V(i)|c\,e^{-t K'}\rangle.
\label{overlapphiT}
\end{align}
To extract the dependence of $t$,
let us consider the operator
\begin{align}
 {\cal L}'_0&=\{Q_{\Psi_0},\,{\cal B}_0\},
\end{align}
where $Q_{\Psi_0}$ is the deformed BRST operator (\ref{deformedBRSTm})
and ${\cal B}_0$ is an anti-ghost operator defined in
Ref.~\cite{Schnabl:2005gv}. As in the case of superstring field theories
\cite{Inatomi:2012nd,Erler:2010pr}, we can derive the equations
\begin{align}
 \frac{1}{2}({\cal L}'_0-{{\cal L}'_0}^\dagger)c&= -c,
\ \ \ 
 \frac{1}{2}({\cal L}'_0-{{\cal L}'_0}^\dagger)K'= K'.
\end{align}
Then, we find 
\begin{align}
 c\,e^{-t\,K'}&=\left(\frac{t}{\kappa}\right)^{1+\frac{1}{2}
({\cal L}'_0-{{\cal L}'_0}^\dagger)}\,c\,e^{-\kappa\,K'},
\label{scaletrans}
\end{align}
where $\kappa$ is a constant. 

Noting that we have \cite{Inatomi:2012nd,Kishimoto:2008zj}
\begin{align}
 \langle I | V(i) ({\cal B}_0-{\cal B}_0^\dagger)&=0,
\ \ \ 
 \langle I | V(i) ({\cal L}'_0-{{\cal L}'_0}^\dagger)=0,
\end{align}
we can rewrite the overlap by using  (\ref{scaletrans}) and perform the
integration in the overlap:
\begin{align}
 O_V(\Phi_T)&
 =\int_0^\infty dt\,e^{-t}\,\frac{t}{\kappa}
\langle
 I|V(i)|c\,e^{-\kappa K'}\rangle
=\frac{1}{\kappa}
\langle
 I|V(i)|c\,e^{-\kappa K'}\rangle.
\end{align}
Then, by following Refs.~\cite{Inatomi:2012nv,Inatomi:2012nd}, we can
express the overlap in terms of a correlation function on
a cylinder of circumference $\pi \kappa/2$:
\begin{align}
O_V(\Phi_T)&=
\frac{2 e^{-\frac{\pi \kappa}{2}{\cal C}}}{\pi \kappa}
\left< \,V(i\infty)\,c(0)\,\exp\left(
\int_0^{\frac{\pi\kappa}{2}} dx\,{\cal J}(x)\right)\,
\right>_{C_{\frac{\pi \kappa}{2}}},
\end{align}
where ${\cal J}(x)$ and ${\cal C}$ are defined by (\ref{calJt}) and
(\ref{calCdef}), respectively. Setting $\kappa=2$ and using the
periodicity along the $x$ direction, we can derive the final expression
of the overlap as in (\ref{correlfunc}).


\end{document}